\documentclass[prl,twocolumn]{revtex4}
\usepackage{epsfig}
\usepackage{bm}
\usepackage{amsmath}

\begin{document}

\title{Dissipative invariants in MHD turbulence}

\author{A. Bershadskii}

\affiliation{
ICAR, P.O. Box 31155, Jerusalem 91000, Israel
}

\begin{abstract}

 Role of the dissipative invariants in isotropic homogeneous MHD turbulence has been studied using the distributed chaos approach and results of direct numerical simulations. Effects of weak imposed mean magnetic field and magnetic fields in undisturbed solar photosphere are also briefly discussed in this context.

\end{abstract}

\maketitle

\section{Introduction}

   The classical inertial ranges with scaling laws (present in abundance, for instance, in the solar wind observations  \cite{bc}-\cite{clv}) are rarely observed in the DNS and laboratory MHD experiments, especially for the quasi-isotropic cases. The obvious reason for this is the strong dissipative effects. 

  Dissipative invariants for the Navier-Stokes equations were introduced by Loitsianskii \cite{my}, Birkhoff \cite{bir} and Saffman \cite{saf}. Chandrasekhar expended this approach on the (resistive) magnetohydrodynamics \cite{cha}. Real area of applicability of these invariants was restricted by dynamics of the mean values and low wavenumbers. Role of the dissipative invariants for other parts of the energy spectra of MHD turbulence is still unclear. Meanwhile these invariants are related to the conservation laws of momentum \cite{saf} and angular momentum \cite{my}, and due to the Noether's theorem to the fundamental spatial symmetries: translational (homogeneity) and rotational (isotropy), respectively. Therefore, one can expect that these invariants should play certain role for other parts of the energy spectrum as well.    \\
  
   The main tool for studying the dissipative invariants is the K\'{a}rm\'{a}n-Howarth equation. For freely decaying isotropic homogeneous MHD turbulence Chandrasekhar obtained the K\'{a}rm\'{a}n-Howarth equation in the form:
$$
\frac{\partial\langle v_Lv_L'\rangle}{\partial t}=\frac{1}{r^4}\frac{\partial}{\partial r} r^4 [\langle v_L^2v_L'\rangle-\langle B_L^2v_l'\rangle]+\frac{2\nu}{r^4}\frac{\partial}{\partial r}r^4{\frac{\partial\langle v_Lv_L'\rangle}{\partial r}}  \eqno{(1)}
$$  
where ${\bf v} = {\bf v} ({\bf x},t)$ is the velocity field,  ${\bf v}' = {\bf v} ({\bf x}+{\bf r}, t)$, and ${\bf B}$ is magnetic filed scaled by $(4\pi \rho)^{1/2}$ ($\rho$ is constant density of the medium). The subscript $_L$ denotes projections of ${\bf v}$ and ${\bf B}$ on ${\bf r}$ (longitudinal, i.e. $v_L = {\bf v} \cdot {\bf r}/r$), $\langle... \rangle$ denotes an ensemble average. Multiplying both sides of the Eq. (1) by $r^4$ and integrating on $r$ from 0 to $R$ one obtains
$$
{\frac{\partial\int\limits_{0}^{R} r^4\langle v_Lv_L'\rangle dr}{\partial t}}= R^4\left( \langle v_L^2v_L'\rangle-\langle B_L^2v_l'\rangle\left.
+2\nu{\frac{\partial\langle v_Lv_L'\rangle}{\partial r}}\right)\right\arrowvert_{r=R} 
$$
if 
$$
\left( \langle v_L^2v_L'\rangle-\langle B_L^2v_l'\rangle \left.
+2\nu{\frac{\partial\langle v_Lv_L'\rangle}{\partial r}}\right)\right\arrowvert_{r=R} 
$$
is approaching to zero fast enough as $R \rightarrow \infty$, then 
$$
\lim_{r\rightarrow R} {\frac{\partial\int\limits_{0}^{R} r^4\langle v_Lv_L'\rangle dr}{\partial t}} =0   \eqno{(2)}
$$
 It means that the dissipative Loitsianskii invariant
$$
\mathcal{L}_v=  \int r^2 \langle v_Lv_L' \rangle d{\bf r} =\rm{constant} \eqno{(3)}
$$
holds also for the MHD case. \\

  In an analogous way (see also below) Chandrasekhar obtained a specific cross-correlation invariant of the Loitsianskii-type for this case:
$$
\mathcal{L}_{cr}=  \int r^2 \langle v_LB_L' \rangle d{\bf r} =\rm{constant}. \eqno{(4)}
$$

\section{More dissipative MHD invariants}   
   
   Let us now consider a K\'{a}rm\'{a}n-Howarth equation in terms of Els\"{a}sser variables \cite{pp}: ${\bf z}^{\pm}={\bf v}\pm {\bf b}$, where ${\bf b} = {\bf B}/\sqrt{\mu_0\rho}$ has the same dimension as the velocity field (the Alfv\'enic units) and ${\bf B}$ is the magnetic field. The K\'{a}rm\'{a}n-Howarth equation can be obtained in following form \cite{pp} (see also below for another form of K\'{a}rm\'{a}n-Howarth equation)
$$
\frac{\partial\langle z_L^{\pm}z_L^{\pm '}\rangle}{\partial t} = \left(\frac{\partial}{\partial r}+\frac{4}{r}\right)C^{\pm}_{LLL}(r)+2\left(\frac{\partial^2}{\partial r^2} 
+ \frac{4}{r} \right) D_{LL}  \eqno{(5)}
$$
where  $D_{LL}(r) = [\nu_+\langle z_L^{\pm}z_L^{\pm '}\rangle+\nu_-\langle z_L^{\pm}z_L^{\mp '}\rangle]$, $C^{\pm}_{LLL}(r)=\langle z_L^{\pm}z_L^{\mp}z_L^{\pm '} \rangle$, $\nu_{\pm} = \nu \pm \eta$, $\nu$ is the viscosity and $\eta$ is the magnetic diffusivity.

  Multiplying both sides of the Eq. (5) by $r^4$ and integrating on $r$ from 0 to $R$ one obtains
$$
{\frac{\partial\int\limits_{0}^{R} r^4\langle z_L^{\pm}z_L^{\pm '}\rangle dr}{\partial t}}=R^4C^{\pm}_{LLL}(R)\left. +2 R^4\frac{\partial D_{LL}}{\partial r}\right\arrowvert_{r=R}  \eqno{(6)}
$$
If
$$
C^{\pm}_{LLL}(R)\left.~~~~\rm{and}~~~~\frac{\partial D_{LL}}{\partial r}\right\arrowvert_{r=R}
$$
are approaching to zero fast enough as $R \rightarrow \infty$, then 
$$
\lim_{r\rightarrow R} {\frac{\partial\int\limits_{0}^{R} r^4\langle z_L^{\pm}z_L^{\pm '}\rangle dr}{\partial t}} =0   \eqno{(7)}
$$
 It means that 
$$
\int r^2 \langle z_L^{\pm}z_L^{\pm '}\rangle d{\bf r} = \rm{constant}  \eqno{(8)}
$$
or, in the terms of ${\bf v}$ and ${\bf b}$ (or ${\bf B}$),
$$
 \int r^2[\langle v_Lv_L'\rangle + \langle b_L  b_L'\rangle]~ d{\bf r} = \rm{constant}  \eqno{(9)}
$$ 
and
$$
 \int r^2 \langle v_LB_L'\rangle~ d{\bf r} = \rm{constant}  \eqno{(10)}
$$ 
The Eq. (10) is the same as Eq. (4). Taking into account the Eq. (3) we obtain from the Eq. (9)
$$
\mathcal{L}_B = \int r^2 \langle B_LB_L' \rangle d{\bf r} =\rm{constant} \eqno{(11)}
$$

   The dissipative invariants $\mathcal{L}_v$, $\mathcal{L}_{cr}$ and $\mathcal{L}_B$ Eqs.(3-4) and (11) are of the Loitsianskii type (related to conservation of angular momentum or to the spatial rotational symmetry - spatial isotropy). In hydrodynamic there is also the Birkhoff-Saffman dissipative invariant \cite{saf} (related to conservation of momentum or to the spatial translational symmetry - homogeneity). In order to generalize the Birkhoff-Saffman invariant on magnetohydrodynamics let us consider the K\'{a}rm\'{a}n-Howarth equation in following form \cite{wan}
$$
\frac{\partial C^{\pm}(r)}{\partial t} = -2 \left(\frac{\partial }{\partial r}+\frac{2}{r}\right)Q^{\pm}(r)+2\nu\left(\frac{\partial^2}{\partial r^2} 
+ \frac{2}{r} \frac{\partial } {\partial r}\right) C^{\pm} (r) \eqno{(12)}
$$
where $C^{\pm} (r) = \langle z_i^{\pm}z_i^{\pm '}\rangle$, $\langle z_k^{\mp '}z_i^{\pm}z_i^{\pm '} \rangle = Q^{\pm} (r) r_k/r$ and $\nu=\eta$ for simplicity.
Multiplying both sides of the Eq. (12) by $r^2$ and integrating on $r$ from 0 to $R$ one obtains
$$
{\frac{\partial\int\limits_{0}^{R} r^2\langle z_i^{\pm}z_i^{\pm '}\rangle dr}{\partial t}}=-2R^2 Q^{\pm} (R)\left. +2\nu R^2\frac{\partial \langle z_i^{\pm}z_i^{\pm '}\rangle}{\partial r}\right\arrowvert_{r=R}  \eqno{(13)}
$$
If
$$
Q^{\pm} (R)\left.~~~~\rm{and}~~~~\frac{\partial \langle z_i^{\pm}z_i^{\pm '}\rangle}{\partial r}\right\arrowvert_{r=R}
$$
are approaching to zero fast enough as $R \rightarrow \infty$, then 
$$
\lim_{r\rightarrow R} {\frac{\partial\int\limits_{0}^{R} r^2\langle z_i^{\pm}z_i^{\pm '}\rangle dr}{\partial t}} =0   \eqno{(14)}
$$
 It means that 
$$
\int \langle z_i^{\pm}z_i^{\pm '}\rangle d{\bf r} = \rm{constant}  \eqno{(15)}
$$
or, in the terms of ${\bf v}$ and ${\bf b}$ (or ${\bf B}$),
$$
\mathcal{S} = \int [\langle {\bf v}\cdot {\bf v}'\rangle + \langle ({\bf b}\cdot {\bf b}'\rangle]~ d{\bf r} = \rm{constant}  \eqno{(16)}
$$ 
and
$$
\mathcal{S}_{cr} = \int \langle {\bf v}\cdot {\bf B}'\rangle~ d{\bf r} = \rm{constant}  \eqno{(17)}
$$ 
The integral $\mathcal{S}$ can be considered as a MHD generalization of the Birkhoff-Saffman invariant \cite{saf}. The cross-correlation integral $\mathcal{S}_{cr}$, as its "Loitsianskii's" counterpart  $\mathcal{L}_{cr}$  Eq. (4), has no analogue in hydrodynamic turbulence. \\
\begin{figure} \vspace{-1.8cm}\centering
\epsfig{width=.45\textwidth,file=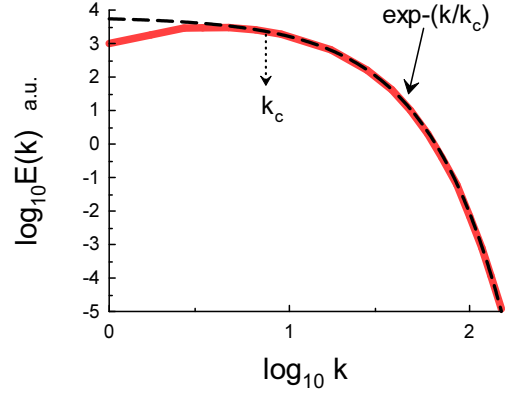} \vspace{-4cm}
\caption{ Magnetic energy spectrum for the DNS with {\it kinetic} energy forcing. } 
\end{figure}
   It should be noted that the dissipative invariants were introduced into hydrodynamics in order to estimate behaviour of the kinetic energy spectrum at small values of wavenumbers $k$ \cite{my},\cite{saf}. For magnetic energy spectrum it can be 
$$
E(k) \propto  \mathcal{L}_B  k^4 \eqno{(18)}
$$
and for kinetic energy spectrum
$$
E(k) \propto  \mathcal{S}  k^2\eqno{(19)}
$$
for small $k$.

\section{Distributed chaos in MHD}
\begin{figure} \vspace{-1.8cm}\centering
\epsfig{width=.45\textwidth,file=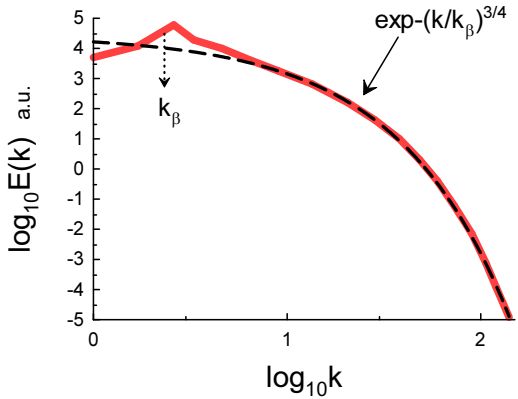} \vspace{-3.8cm}
\caption{ Kinetic energy spectrum for the DNS with {\it kinetic} energy forcing (for the same DNS as in the Fig. 1). } 
\end{figure}
\begin{figure} \vspace{-0.3cm}\centering
\epsfig{width=.45\textwidth,file=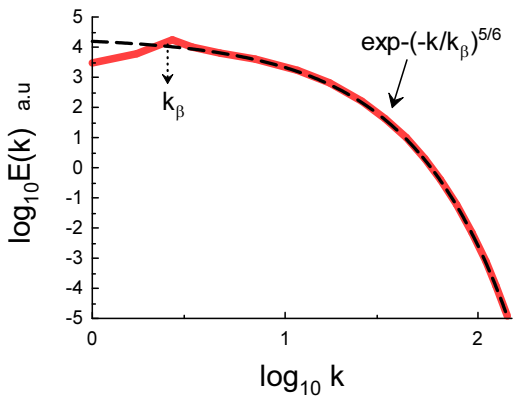} \vspace{-4cm}
\caption{ Magnetic energy spectrum for the DNS with simultaneous forcing of the kinetic energy and cross helicity.} 
\end{figure}
    In the fluids/plasmas dynamics at the onset of turbulence the deterministic chaos is, as a rule, related to spatial exponential spectra \cite{mm},\cite{kds}
$$
E(k) \propto \exp-(k/k_c)  \eqno{(20)}
$$       
where $k_c$ is a constant. For magnetohydrodynamics with the {\it kinetic} energy forcing situation can be, naturally, more complex than in hydrodynamics. In this situation the magnetic energy spectrum can be still exponential while the kinetic energy spectrum already corresponds to a distributed chaos/turbulence state (see Figs. 1-2 and next Section for more detailed description of the direct numerical simulation - the spectral data for the Figs. 1-2 were taken from Fig. 3 of the Ref. \cite{step}). The dashed curve in the Fig. 1 indicates correspondence to the exponential spectrum Eq. (20) in the log-log scales and the dotted arrow indicates position of the $k_c$.  

   When the deterministic chaos is transformed into developed turbulence the parameter $k_c$  can fluctuate and we should use an ensemble averaging
$$
E(k) \propto \int_0^{\infty} P(k_c) \exp -(k/k_c)dk_c, \eqno{(21)}
$$    
introducing a probability density distribution $P(k_c)$. In order to find $P(k_c)$ the dimensional considerations and the dissipative invariants can be used. 

   Stretched exponential is a natural generalization of the ordinary exponential function
$$
E(k) \propto \int_0^{\infty} P(k_c) \exp -(k/k_c)dk_c  \propto \exp-(k/k_{\beta})^{\beta},  \eqno{(22)}
$$   
   The asymptote of the $P(k_c)$ at large $k_c$ can be found from the Eq. (22) \cite{jon}
$$
P(k_c) \propto k_c^{-1 + \beta/[2(1-\beta)]}~\exp(-bk_c^{\beta/(1-\beta)}) \eqno{(23)}
$$  
 
 On the other hand, a relationship between characteristic strength of the magnetic field $B_c$ and $k_c$ can be obtained from the dimensional considerations as 
$$
B_c \propto |\mathcal{L}_{B}|^{1/2} k_c^{5/2}  \eqno{(24)}
$$
or, generally, 
$$ 
B_c \propto  k_c^{\alpha}  \eqno{(25)}
$$  
  In the case of normally distributed $B_c$ (with zero mean) we obtain from the Eqs. (23) and (25) relationship between $\beta$ and $\alpha$
$$
\beta = \frac{2\alpha}{1+2\alpha}  \eqno{(26)}
$$

   Substitution of the $\alpha =5/2$ from the Eq. (24), in particular, gives 
$$
E(k) \propto \exp-(k/k_{\beta})^{5/6}  \eqno{(27)}
$$
for  magnetic energy spectrum.\\

  Analogously we can find kinetic energy spectrum corresponding to domination of the MHD Birkhoff-Saffman invariant $\mathcal{S}$ (the Eq. (16) in the Alfv\'enic units) in the velocity field. Relationship between characteristic velocity $v_c$ and $k_c$ can be obtained from the dimensional considerations as 
$$
v_c \propto |\mathcal{S}|^{1/2} k_c^{3/2}  \eqno{(28)}
$$ 
In the case of normally distributed $v_c$ (with zero mean) we obtain from Eqs. (26) and (28)
$$
E(k) \propto \exp-(k/k_{\beta})^{3/4}.  \eqno{(29)}
$$
for the kinetic energy spectrum.

\section{Direct numerical simulations}
 
\begin{figure} \vspace{-1.8cm}\centering
\epsfig{width=.45\textwidth,file=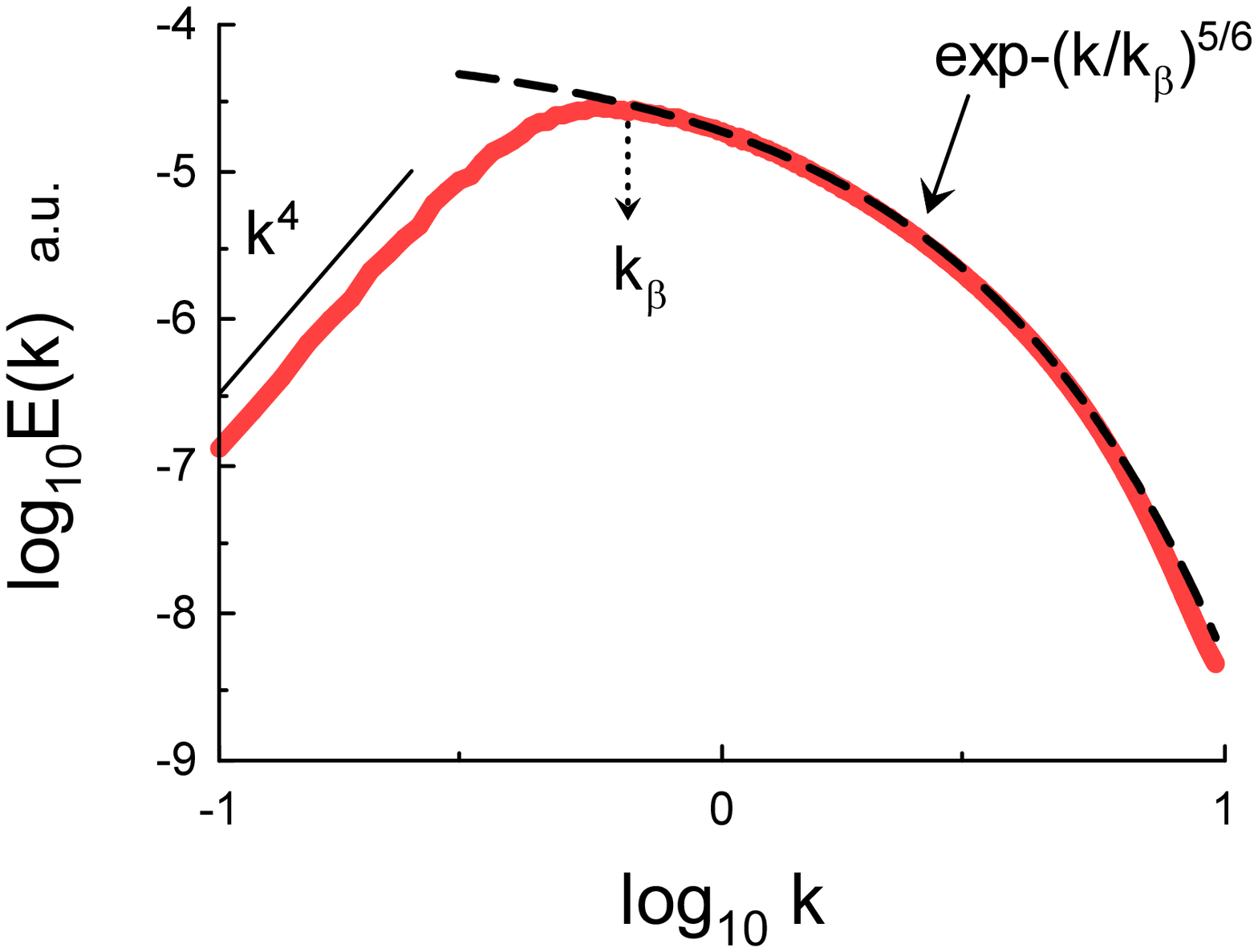} \vspace{-3.7cm}
\caption{ Magnetic energy spectrum for the freely decaying MHD turbulence at $t/\tau_0 \simeq200$. } 
\end{figure}
\begin{figure} \vspace{-0.5cm}\centering
\epsfig{width=.45\textwidth,file=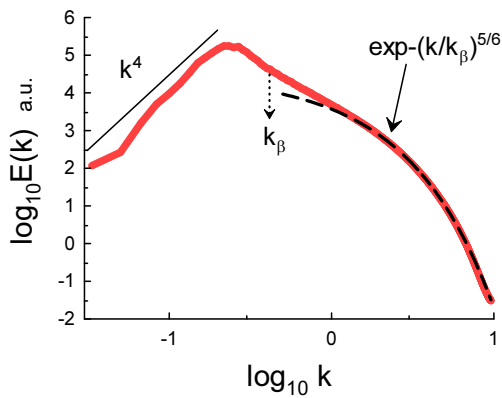} \vspace{-4.2cm}
\caption{ Magnetic energy spectrum for the freely decaying MHD turbulence at $t/\tau_0 \simeq 2625$. } 
\end{figure}

  When in the above mentioned DNS \cite{step} a cross helicity forcing was added to the kinetic energy forcing the magnetic field was stirred and forced into the distributed chaos regime with magnetic energy spectrum corresponding to the Eq. (27), as one can see from the Fig. 3 (cf. Fig. 1, the spectral data for Fig. 3 were taken from the Fig. 3 of the Ref. \cite{step} ). The kinetic energy spectrum in this case was very similar to that shown in the Fig. 2 (i.e. without the cross helicity forcing) and only value of $k_{\beta}$ was slightly changed, whereas the decaying part of the spectrum was still following the distributed chaos stretched exponential Eq. (29) (the dashed curve in the Fig. 2). 
  
    The position of the $k_{\beta}$ in the Fig. 3 indicates that the entire distributed chaos was tuned to the forcing, which was applied in the range of large-scales $2 < k <3$ in this case.
    
    In this DNS a statistically stationary isotropic and homogeneous MHD turbulence was simulated in a cubic volume with periodic boundary conditions using MHD equations for incompressible fluid in the Alfv\'enic units:
$$
 \frac{\partial {\bf v}}{\partial t} = - {\bf v} \cdot \nabla {\bf v} 
    -\frac{1}{\rho} \nabla {\cal P} - [{\bf b} \times (\nabla \times {\bf b})] + \nu \nabla^2  {\bf v} + {\bf f}_v \eqno{(30)}
$$
$$
\frac{\partial {\bf b}}{\partial t} = \nabla \times ( {\bf v} \times
    {\bf b}) +\eta \nabla^2 {\bf b}    \eqno{(31)} 
$$
$$ 
\nabla \cdot {\bf v}, ~~~~~~~~~~~\nabla \cdot {\bf b},  \eqno{(32)}
$$
For the DNS $\nu =0.008$, $\eta = 0.004$ and the large-scale forcing term is 
$$
{\bf f}_v = \frac{ {\bf v}~}{|{\bf v}|^2} +\frac{\gamma {\bf b}}{|{\bf v}||{\bf b}|} .  \eqno{(33)}
$$
 In the case when the only kinetic energy forcing was applied the parameter $\gamma =0$, whereas in the case when the cross helicity forcing was added $\gamma =1$. The averaged magnetic helicity $ \langle {\bf A} \cdot {\bf B} \rangle =0$ in this DNS (where ${\bf B} = \nabla \times {\bf A}$).\\ 
  
  Another interesting direct numerical simulation of freely decaying isotropic and homogeneous MHD turbulence (with a considerable initial kinetic helicity) was performed in recent Ref. \cite{bran}. In this DNS the initial magnetic field was weak and a weakly compressible gas was considered (mainly for the computational efficiency, because the compressibility effects were unimportant for the Mach numbers $\sim 0.1$), $\nu = \eta$. The initial velocity field was taken to be solenoidal with normally distributed fluctuations and the periodic boundary conditions were applied. Initial magnetic helicity (in the considered here case) was equal to zero and at the early times of the decay was negligible. 
  
\begin{figure} \vspace{-1.35cm}\centering
\epsfig{width=.45\textwidth,file=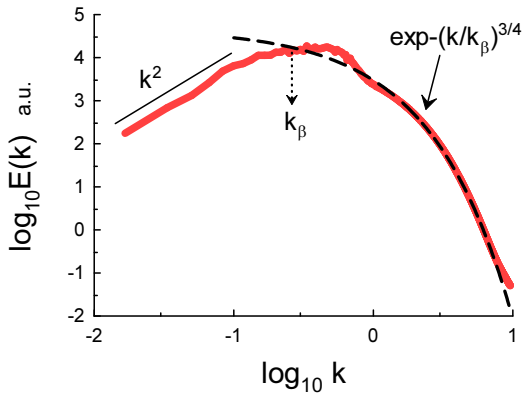} \vspace{-3.95cm}
\caption{ Kinetic energy spectrum for the freely decaying MHD turbulence at $t/\tau_0 \simeq 2625$. } 
\end{figure}
  
\begin{figure} \vspace{-0.5cm}\centering
\epsfig{width=.45\textwidth,file=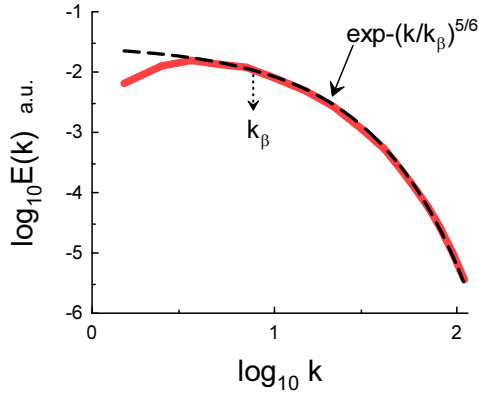} \vspace{-4.23cm}
\caption{Magnetic energy spectrum for forced MHD turbulence with a weak imposed magnetic field.} 
\end{figure}
    
   Figure 4 shows, in the log-log scales, magnetic energy spectrum at $t/\tau_0 \simeq200$ (the early times of the decay). The spectral data are available at the site  \cite{bran2} ($\tau_0$ is the initial turnover time, see for the more detail the Ref. \cite{bran}). The dashed curve indicates correspondence to the Eq. (27). The dotted arrow indicates position of $k_\beta$. From this position one can assume that in this case the entire distributed chaos was tuned to the large-scale coherent structures. The straight line indicates corresponding spectral behaviour at small $k$ (see the Eq.(18)). It should be emphasized that $\mathcal{L}_B$ was the dominating dissipative invariant both for the small (the Eq. (18)) and for the large  values of $k$ (the Eqs. (24) and (27)) for the magnetic field in this case. 
   
   Figure 5 shows magnetic energy spectrum at time of decay $t/\tau_0 \simeq 2625$ (cf. Fig. 4 and the comments to it). Figure 6 shows kinetic energy spectrum at the same time $t/\tau_0 \simeq 2625$ (cf. Fig. 2 for the forced MHD turbulence). The dashed curve in the Fig. 6 indicates correspondence to the Eq. (29). The straight line indicates corresponding spectral behaviour at small $k$ (now it is the Eq.(19)). It should be emphasized that $\mathcal{S}$ was the dominating invariant both for the small (the Eq. (19)) and for the large values of $k$ (the Eqs. (28-29)) for the velocity field in this case.\\

   In paper Ref. \cite{cho} effects of imposed mean magnetic field on isotropic homogeneous incompressible MHD turbulence were studied using direct numerical simulations in a periodic box. A large scale forcing was applied and $\nu = \eta$. For a weak imposed magnetic field ($b_0 < 0.2$ in the Alfv\'enic units) 
 the turbulence can be still considered as statistically isotropic in this DNS \cite{cho}. On the other hand, it is mentioned in the Ref. \cite{wan} that imposed mean magnetic field does not appear explicitly in the K\'{a}rm\'{a}n-Howarth equation, and the problem with its application to the case is violation of the isotropy by the mean magnetic field.  Therefore, one can expect that the above consideration can be still applied in the case of the weak magnetic field. Figure 7 shows magnetic energy spectrum for $b_0 =0.1$. The spectral data were taken from Fig. 11 of the Ref. \cite{cho}. The dashed curve indicates correspondence to the Eq. (27) (cf. Figs. 4,5).
 
\section{Magnetic fields in undisturbed solar photosphere} 

\begin{figure} \vspace{-0.4cm}\centering
\epsfig{width=.45\textwidth,file=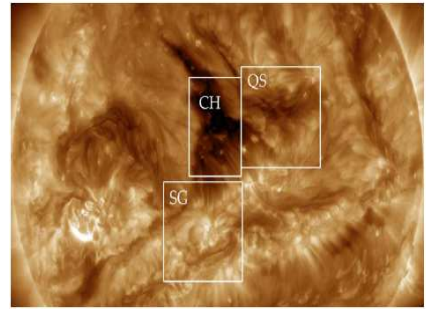} \vspace{-4.93cm}
\caption{Typical coronal hole (CH), quiet Sun (QS) and super-granulation (SG) areas.} 
\end{figure}

 The Sun's surface is mostly undisturbed. The typical areas of the undisturbed Sun surface are the coronal holes (CH), the quite Sun areas, and the super-granulation areas. The magnetic fields in the undisturbed areas are weak and their origin is still unclear. They are especially weak in the coronal holes, somewhat stronger in the areas of the quiet Sun and have an additional strength in the super-granulation areas (which presumably comes from the previous decaying disturbances).\\
 
  Application of the previous consideration to these magnetic fields is complicated by the effects introduced by the Sun's rotation and the thermal convection in the photosphere and beneath. However, the  Loitsianskii-like  invariants can be generalised on these phenomena (see, for instance, Ref. \cite{dav}, and for application of the distributed chaos approach to the Sun's magnetic fields Ref. \cite{b2}). \\
  
\begin{figure} \vspace{-1.5cm}\centering
\epsfig{width=.45\textwidth,file=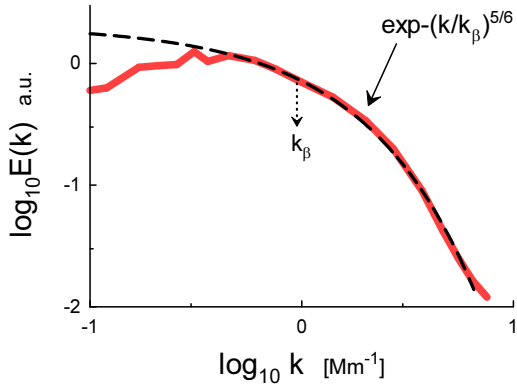} \vspace{-4.23cm}
\caption{Magnetic energy spectrum for the coronal hole area.} 
\end{figure}
\begin{figure} \vspace{-0.5cm}\centering
\epsfig{width=.45\textwidth,file=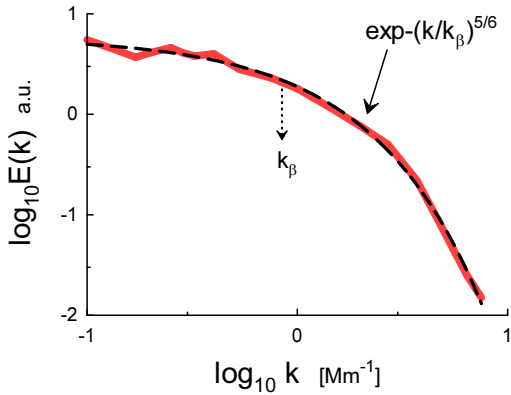} \vspace{-4.23cm}
\caption{Magnetic energy spectrum for the quiet Sun area.} 
\end{figure}
\begin{figure} \vspace{-0.5cm}\centering
\epsfig{width=.45\textwidth,file=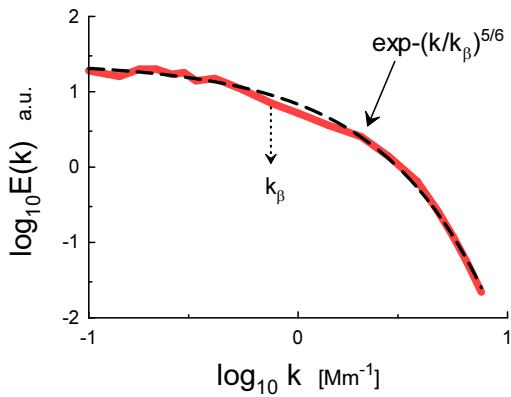} \vspace{-4.23cm}
\caption{Magnetic energy spectrum for the super-granulation area.} 
\end{figure}

  Figure 8 (adapted from the Fig. 3 of the recent Ref. \cite{ak}) shows the coronal hole (CH), quiet Sun (QS) and super-granulation (SG) areas analysed in the Ref. \cite{ak} using the data obtained in the undisturbed photosphere with the Helioseismic and Magnetic Imager on board Solar Dynamic Observatory.\\
  
  Figures 9-11 shows power spectra of the magnetic fields for the coronal hole (CH), quiet Sun (QS) and super-granulation (SG) areas correspondingly (the spectral data were taken from Fig. 4 of the Ref. \cite{ak}, and the measurements were produced on 2015 March 10). The spectra were computed using the method described in the Ref. \cite{abr}. The dashed curves indicate correspondence to the Eq. (27) and the dotted arrows indicate position of the $k_{\beta}$.

\section{Acknowledgement}

  I thank A. Brandenburg for sharing his data and additional information.

\end{document}